\documentclass[reprint,article,twocolumn,
superscriptaddress,
amsmath,amssymb,
aps,
pra,
floatfix,
]{revtex4-1}

\usepackage{apjfonts}
\usepackage{graphicx}
\usepackage[colorlinks=true,citecolor=blue,linkcolor=blue,urlcolor=blue]{hyperref}
\usepackage{siunitx}
\usepackage{float}

\begin{document}

\title{Observation of strong two-electron--one-photon transitions in few-electron ions}

\author{M. Togawa}
\email[]{moto.togawa@mpi-hd.mpg.de}
\affiliation{Max-Planck-Institut f\"ur Kernphysik, Saupfercheckweg 1, 69117 Heidelberg, Germany} 

\author{S. K\"uhn}
\email[]{steffen.kuehn@mpi-hd.mpg.de}
\affiliation{Max-Planck-Institut f\"ur Kernphysik, Saupfercheckweg 1, 69117 Heidelberg, Germany}
\affiliation{Heidelberg Graduate School of Fundamental Physics, Ruprecht-Karls-Universit\"at Heidelberg, Im Neuenheimer Feld 226, 69120 Heidelberg, Germany}

\author{C. Shah}
\email[]{chintan.d.shah@nasa.gov}
\affiliation{NASA Goddard Space Flight Center, 8800 Greenbelt Road, Greenbelt, Maryland 20771, USA}

\affiliation{Max-Planck-Institut f\"ur Kernphysik, Saupfercheckweg 1, 69117 Heidelberg, Germany}

\author{P. Amaro}
\email[]{pdamaro@fct.unl.pt}
\affiliation{Laboratory of Instrumentation, Biomedical Engineering and Radiation Physics (LIBPhys-UNL), Department of Physics, NOVA School of Science and Technology, NOVA University Lisbon, 2829-516 Caparica, Portugal}

\author{R. Stein\-br\"ugge}
\affiliation{Deutsches Elektronen-Sychrotron DESY, Notkestra{\ss}e 85, 22607 Hamburg, Germany}

\author{J. Stierhof}
\affiliation{Dr.~Karl Remeis-Sternwarte and Erlangen Centre for Astroparticle Physics, Sternwartstra{\ss}e 7, 96049 Bamberg, Germany}

\author{N. Hell}
\affiliation{Lawrence Livermore National Laboratory, 7000 East Avenue, Livermore, California 94550, USA}

\author{M. Rosner}
\affiliation{Max-Planck-Institut f\"ur Kernphysik, Saupfercheckweg 1, 69117 Heidelberg, Germany}
\affiliation{Heidelberg Graduate School of Fundamental Physics, Ruprecht-Karls-Universit\"at Heidelberg,
Im Neuenheimer Feld 226, 69120 Heidelberg, Germany}

\author{K. Fujii}
\affiliation{Department of Mechanical Engineering and Science, Graduate School of Engineering, Kyoto University, Kyoto 615-8540, Japan}

\author{M. Bissinger}
\affiliation{Dr.~Karl Remeis-Sternwarte and Erlangen Centre for Astroparticle Physics, Sternwartstra{\ss}e 7, 96049 Bamberg, Germany}

\author{R. Ballhausen}
\affiliation{Dr.~Karl Remeis-Sternwarte and Erlangen Centre for Astroparticle Physics, Sternwartstra{\ss}e 7, 96049 Bamberg, Germany}

\author{M. Hoesch}
\affiliation{Deutsches Elektronen-Sychrotron DESY, Notkestra{\ss}e 85, 22607 Hamburg, Germany}

\author{J. Seltmann}
\affiliation{Deutsches Elektronen-Sychrotron DESY, Notkestra{\ss}e 85, 22607 Hamburg, Germany}

\author{S. Park}
\affiliation{Ulsan National Institute of Science and Technology, 50 UNIST-gil, Ulsan 44919, South Korea}

\author{F. Grilo}
\affiliation{Laboratory of Instrumentation, Biomedical Engineering and Radiation Physics (LIBPhys-UNL), Department of Physics, NOVA School of Science and Technology, NOVA University Lisbon, 2829-516 Caparica, Portugal}

\author{F. S. Porter}
\affiliation{NASA Goddard Space Flight Center, 8800 Greenbelt Road, Greenbelt, Maryland 20771, USA}

\author{J. P. Santos}
\affiliation{Laboratory of Instrumentation, Biomedical Engineering and Radiation Physics (LIBPhys-UNL), Department of Physics, NOVA School of Science and Technology, NOVA University Lisbon, 2829-516 Caparica, Portugal}

\author{M. Chung}
\affiliation{Ulsan National Institute of Science and Technology, 50 UNIST-gil, Ulsan 44919, South Korea}

\author{T. St\"ohlker}
\affiliation{Institut f\"ur Optik und Quantenelektronik, Friedrich-Schiller-Universit\"at Jena, F\"urstengraben 1, 07743 Jena, Germany} 
\affiliation{GSI Helmholtzzentrum f\"ur Schwerionenforschung, Planckstra{\ss}e 1, 64291 Darmstadt, Germany}
\affiliation{Helmholtz-Institut Jena, Fr\"obelstieg 3, 07743 Jena, Germany} 

\author{J. Wilms}
\affiliation{Dr.~Karl Remeis-Sternwarte and Erlangen Centre for Astroparticle Physics, Sternwartstra{\ss}e 7, 96049 Bamberg, Germany}

\author{T. Pfeifer}
\affiliation{Max-Planck-Institut f\"ur Kernphysik, Saupfercheckweg 1, 69117 Heidelberg, Germany}

\author{G. V. Brown}
\affiliation{Lawrence Livermore National Laboratory, 7000 East Avenue, Livermore, California 94550, USA}

\author{M. A. Leutenegger}
\affiliation{NASA Goddard Space Flight Center, 8800 Greenbelt Road, Greenbelt, Maryland 20771, USA}

\author{S. Bernitt}
\affiliation{Max-Planck-Institut f\"ur Kernphysik, Saupfercheckweg 1, 69117 Heidelberg, Germany} 
\affiliation{GSI Helmholtzzentrum f\"ur Schwerionenforschung, Planckstra{\ss}e 1, 64291 Darmstadt, Germany}
\affiliation{Institut f\"ur Optik und Quantenelektronik, Friedrich-Schiller-Universit\"at Jena, F\"urstengraben 1, 07743 Jena, Germany} 
\affiliation{Helmholtz-Institut Jena, Fr\"obelstieg 3, 07743 Jena, Germany}

\author{J. R. {Crespo L\'opez-Urrutia}}
\email[]{crespojr@mpi-hd.mpg.de}
\affiliation{Max-Planck-Institut f\"ur Kernphysik, Saupfercheckweg 1, 69117 Heidelberg, Germany}

\date{Received 9 March 2020; revised 24 September 2020; accepted 28 September 2020; published 25 November 2020}                    

\begin{abstract}
We resonantly excite the $K$ series of O$^{5+}$ and O$^{6+}$ up to principal quantum number $n=11$ with monochromatic x rays, producing $K$-shell holes, and observe their relaxation by soft-x-ray emission. Some photoabsorption resonances of O$^{5+}$ reveal strong two-electron--one-photon (TEOP) transitions. We find that for the $[(1s\,2s)_1\,5p_{3/2}]_{3/2;1/2}$ states, TEOP relaxation is by far stronger than the radiative decay and competes with the usually much faster Auger decay path. This enhanced TEOP decay arises from a strong correlation with the near-degenerate upper states $[(1s\,2p_{3/2})_1\,4s]_{3/2;1/2}$ of a Li-like satellite blend of the He-like $K\alpha$ transition. Even in three-electron systems, TEOP transitions can play a dominant role, and the present results should guide further research on the ubiquitous and abundant many-electron ions where electronic energy degeneracies are far more common and configuration mixing is stronger.  
\\
\,\,\,  \\
{DOI: \href{https://link.aps.org/doi/10.1103/PhysRevA.102.052831}{10.1103/PhysRevA.102.052831}}
\end{abstract}

\maketitle

\section{Introduction}

In hot astrophysical plasmas, the most common elements, hydrogen and helium, are fully ionized, and only those with higher nuclear charge can keep some bound electrons, appearing as highly charged ions (HCIs) \cite{Beiersdorfer2003}. The widths, Doppler shifts and relative intensities of their characteristic lines are recorded by x-ray observatories and analyzed for plasma diagnostics, relying not only on tabulated calculations but also on more scarce laboratory data. To fully exploit the data of current and upcoming high-resolution x-ray missions such as \textit{XRISM}~\cite{XRISM} and \textit{Athena}~\cite{ATHENA}, more accurate laboratory tests of the atomic models used in astrophysics are needed \cite{Beiersdorfer2003,Hell2020}. 
Light elements such as carbon, nitrogen, and the oxygen studied here abundantly appear as HCIs over a broad range of temperatures and can thus serve as unique spectroscopic probes of, e.g., the warm-hot intergalactic medium (WHIM), which is critical to a complete census of baryonic matter in the universe~\cite{WHIM,Nicastro2018,Gatuzz2019}. It is important to have knowledge of both the photoabsorption cross sections and the various decay channels that govern the fluorescence yield and the ionization balance in plasmas. After x-ray absorption takes place, the most common relaxation processes are direct radiative decay and autoionization. However, even in few-electron systems, more complex processes and multi-electron transitions also compete with them. Including such mechanisms in models is computationally intensive, and hence laboratory data are needed to guide those efforts~\cite{kallman2007}.

Many-electron processes are intensively studied in both theory and experiment, and there is a plethora of recent examples on various subjects: multiple photodetachment of anions (see e.~g., \cite{Gorczyca2004,Schippers2016,Mueller2018,Perry2020} and references therein); photoionization of atoms and ions \cite{West2001,Kjeldsen2006,Bizau2015,Berrah2004,Gharaibeh2011} near inner-shell absorption edges \cite{Levin2004,Kabachnik2007,Southworth2019,Ma2017}; and higher-order relaxation processes \cite{Dunford2004}. This also applies to ions (see e.~g., \cite{Schippers2016b,Mueller2017}, HCI \cite{Blancard2018,Simon2010,Steinbruegge2015,Bliman_1989,Tawara2002,Kato2006} and their interactions with free-electron lasers \cite{Young2010,Buth2018}. Photorecombination also triggers multi-electronic excitations through resonant dielectronic \cite{Beiersdorfer1992,Nakamura2008,Shah2015}, trielectronic and quadruelectronic processes \cite{Schnell2003,Beilmann2011,ShahPRE2016}. The complexity of interelectronic correlations already within the $L$ shell \cite{drake1999,Bizau2015,Liu2018,Zaytsev2019} forces theoreticians to use approximations with uncertainties that are hard to benchmark in the absence of laboratory data. As an example, the crucial determination of the cosmic abundance and column-density of O$^{5+}$ in the WHIM suffers from large theoretical uncertainties \cite{Behar2001,McLaughlin2016,Gatuzz2019,Mathur2017}.

Here, we report on resonant excitation of the $K$ series of He-like and Li-like oxygen ions between 570 and 750\,eV using monoenergetic soft x rays. We detect their fluorescence-photon yield and energy as a function of the incident photon energy, and observe surprisingly strong and sometimes dominating two-electron--one-photon (TEOP) transitions in Li-like oxygen.

\section{Experimental setup}
\begin{figure}%[t]
\includegraphics[width=0.48\textwidth]{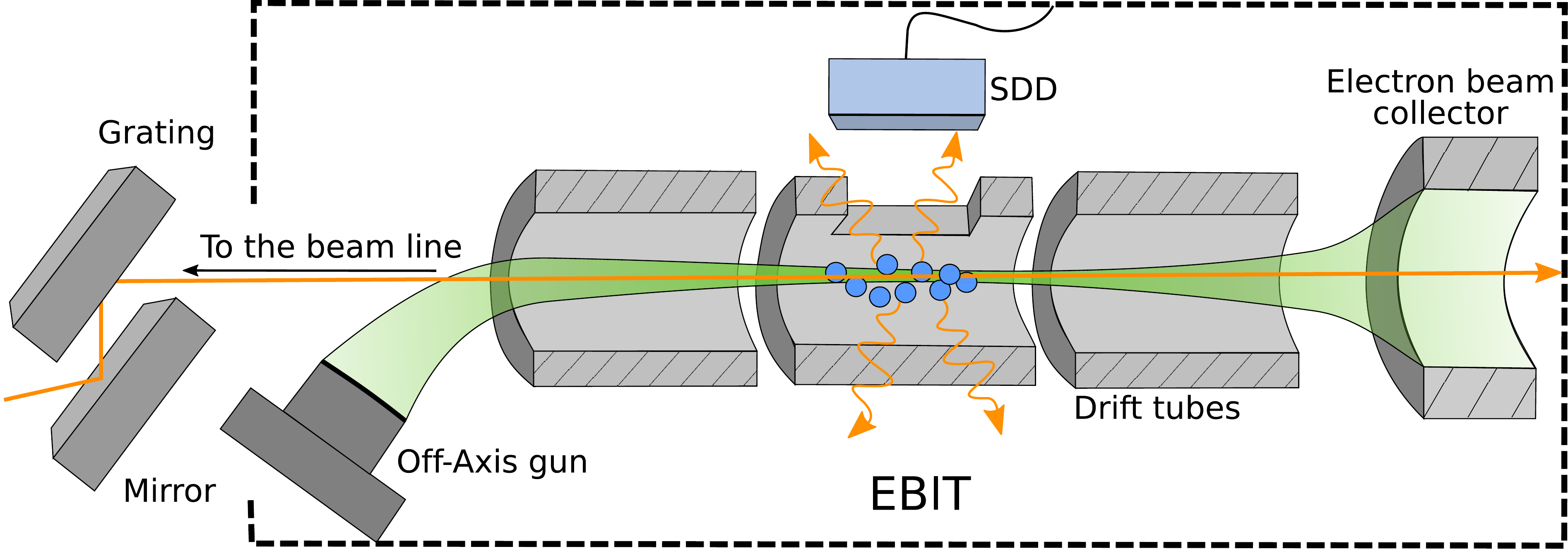}
\begin{center}
\caption{Schematic of PolarX EBIT \cite{Micke2018}. An electron beam from an off-axis gun is focused by a magnetic field and passes through drift tubes, where it generates and traps highly charged ions before reaching a collector electrode. A monochromatic photon beam enters axially and excites the trapped ions. The energies of fluorescence photons are recorded by a silicon-drift detector (SDD), similarly to \cite{Kuehn2020}.}
\label{aufbau2}
 \end{center}
\end{figure}

Our experiment was conducted at the variable-polarization XUV beamline P04 \cite{Viefhaus2013} of the PETRA III synchrotron facility with a portable electron beam ion trap (EBIT), PolarX \cite{Micke2018} (see Fig.~\ref{aufbau2}). Molecular oxygen was injected into the EBIT, dissociated and successively ionized yielding a large He-like O$^{6+}$ population and a small Li-like O$^{5+}$ fraction. These HCI are radially trapped by the electron beam (here 3\,mA, to reduce ion heating), and axially confined within a potential well formed by making the central drift tube slightly more negative than the adjacent ones. With an electron-beam energy of $\sim$200\,eV just above the Li-like ionization threshold, we produce He-like O$^{6+}$, but stay below the excitation threshold of $K\alpha$ or higher $Kn$ series transitions. This ensures a low-background measurement of the $K$ series fluorescence by a silicon-drift detector (SDD) mounted side-on above the central drift tube where the ions are confined.

\begin{figure}%[thb]
\includegraphics[width=0.48\textwidth]{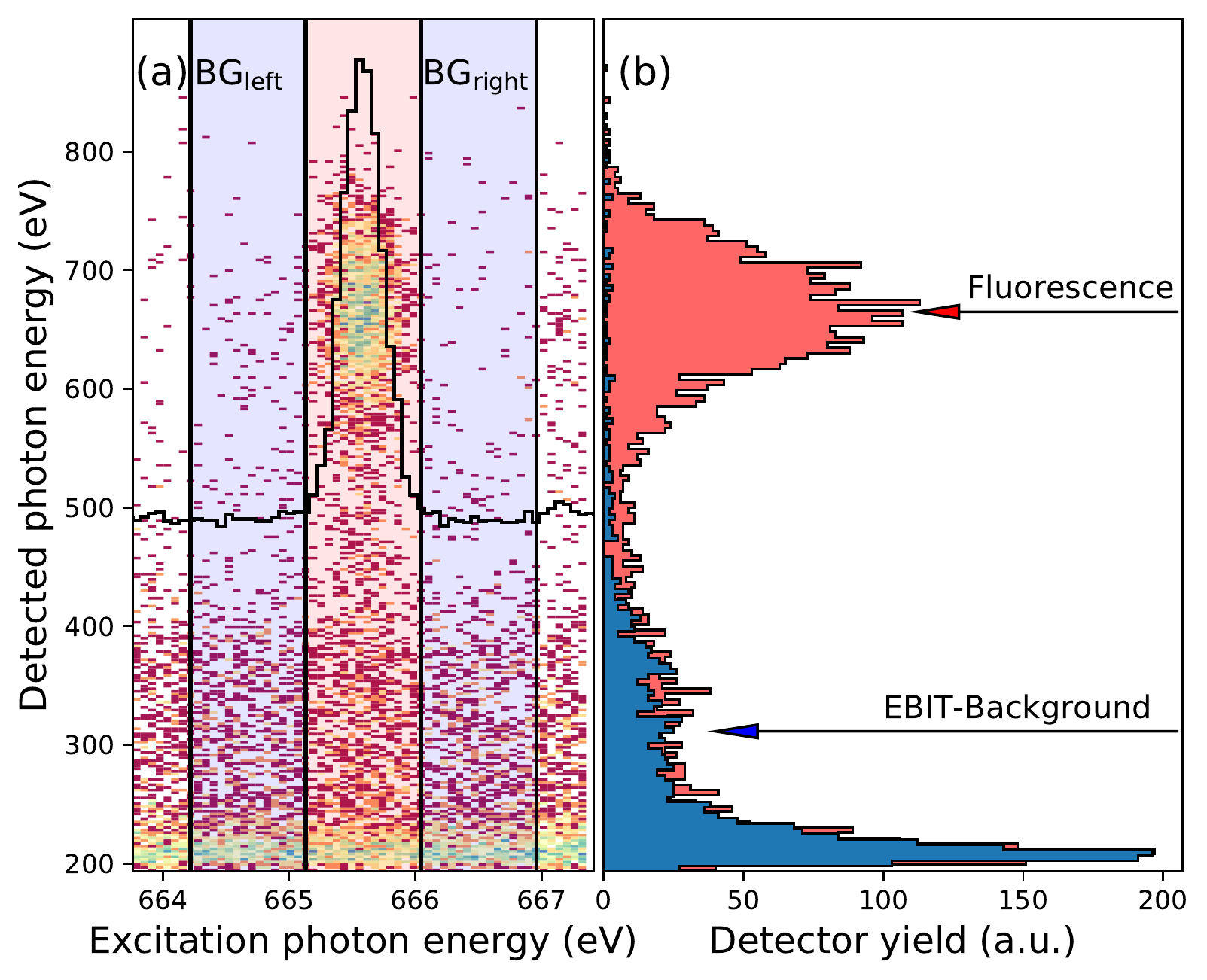}
\caption{a) Fluorescence yield (recorded by the SDD) of He-like O$^{6+}$ $K_\beta$ under excitation with \SI{663}-\SI{668}{\electronvolt} photons ($x$ axis) versus fluorescence-photon energy ($y$ axis), and (black curve) projection of the photon yield onto $x$. b) Projections onto the $y$ axis of on/off resonance slices (red/blue histograms). The off-projection is used for subtraction of electron-induced background and SDD-noise.}
\label{projection}
\end{figure}

\begin{figure*}[t]
\centering \includegraphics[width=0.95\textwidth]{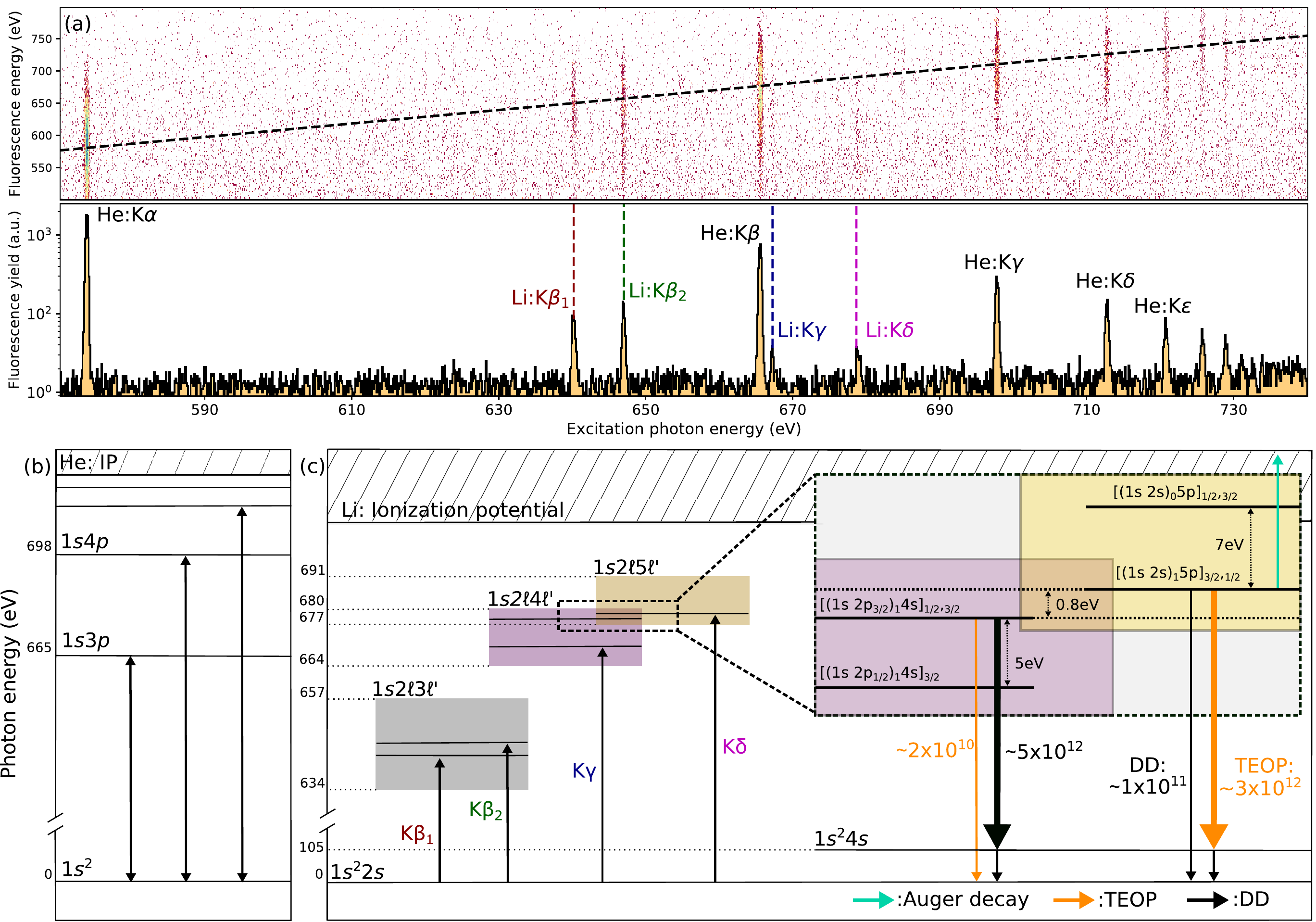}
\caption{
\textbf{a)} Shown on top is a histogram of events registered with the SDD: fluorescence photon energy ($y$ axis) versus excitation photon energy ($x$ axis). The dashed diagonal line indicates elastic channels, with excitation and the fluorescence photon at equal energies. 
Shown on the bottom is fluorescence from resonantly photoexcited He-like O$^{6+}$ and Li-like O$^{5+}$, labeled in black and in color, respectively. 
\textbf{b)} He-like and \textbf{c)} Li-like level diagrams. Configuration mixing of the upper levels of K$\delta$ causes a predominance of the TEOP process over both the elastic direct decay to the ground state and the Auger channels. Rates are in units of s$^{-1}$.
}
\label{fullview}
\end{figure*}

The P04 beamline is equipped with an APPLE-II undulator covering the photon energy range \SI{250} -- \SI{3000}\electronvolt, and a grating monochromator (\SI{1200} lines/mm) providing circularly polarized light at a resolving power of more than \SI{e4}{} \cite{Viefhaus2013}. Expected long-time drifts of the monochromator recommended for this overview measurement a fast-scan mode lasting less than one hour, forcing the use of a wide slit (200\,$\mu$m) for better statistics. This gives us a photon flux on the order of $10^{12}$ photons/s at a moderate resolving power. Nonetheless, we could determine excitation energies with a relative uncertainty of $\Delta E/E\approx 10^{-5}$. For this, we digitize the SDD-energy signal ($y$ axis) for each photon-detection event while continuously scanning the monochromator, i.~e., the incident photon energy ($x$ axis), obtaining a two-dimensional fluorescence histogram (Fig. \ref{fullview}a). To remove background events due to electron recombination, we subtract at each resonance the off-resonance mean count rate from the on-resonance signal (see an example in Fig.~\ref{projection}). Then, we project the region of interest containing the resonance onto both axes, and fit Gaussians with a full widths at half maximum of order of $\SI{350}{\milli\electronvolt}$ ($x$ axis) and order of $\SI{100}{\electronvolt}$ ($y$ axis) to those projections. 

%\section{MEASUREMENT OF PHOTON-EXCITATION ENERGIES} % we should use photoexcitation energies instead. (we did not use photon-excitation term elsewhere in the paper
\section{MEASUREMENT OF PHOTOEXCITATION ENERGIES}
A monochromator scan from $\SI{570}{\electronvolt}$ to $\SI{740}{\electronvolt}$ at $\SI{500}{\milli \electronvolt \per \second}$ resolved the core excited $K$ series of He-like oxygen up to $K\kappa$, as well as several other weaker Li-like resonances (Fig.~\ref{fullview}). 
After determining their centroid positions (see Fig.~3a), we assign to the six transitions $K\alpha$ up to $K\zeta$ (identified on the approximately calibrated $x$ axis for the incident photon-energy in Fig.~\ref{fullview}) energies taken from accurate calculations by \citet{Yerokhin2019} with uncertainties on the order of $\SI{0.5}{\milli \electronvolt}$ and determine the final monochromator-dispersion curve with a linear fit. Its confidence interval is basically dominated by the order of 20-meV statistical uncertainties of the individual transitions in our fast overview scan. By extrapolating the dispersion curve we obtain the excitation energies of the $K\eta$, $K\theta$, $K\iota$ and $K\kappa$ transitions (see Table~\ref{tab1}).
Using these data points, we are able to determine the ionization potential of O$^{6+}$. We use a quantum-defect model based on the Rydberg formula, with the Rydberg energy $E_\text{R}$, the effective nuclear charge $Z_\text{eff}$ and the quantum defect $\delta_{n,l}$ for principal $n$ and orbital $l$ quantum numbers, respectively:
%
%\begin{eqnarray}\nonumber
%K_n & = &  Z_\text{eff}^2E_\text{R}\left[{(1-\delta_{1,s})^{-2}}-{(n-\delta_{n,l})^{-2}}\right]\\ 
%     & = & E_\text{IP}-Z_\text{eff}^2E_\text{R}{(n-\delta_{n,l})^{-2}}.
%\label{rydberg}
%\end{eqnarray}
%
\begin{eqnarray}\nonumber
		K_n & = &  Z_\text{eff}^2E_\text{R}\left[{(1-\delta_{1,s})^{-2}}-{(n-\delta_{n,l})^{-2}}\right]\\ 
     			& \equiv & E_\text{IP}-Z_\text{eff}^2E_\text{R}{(n-\delta)^{-2}}.
			\label{rydberg}
		\end{eqnarray}

\begin{figure}[b]
	\begin{center}
		\includegraphics[width=0.5\textwidth]{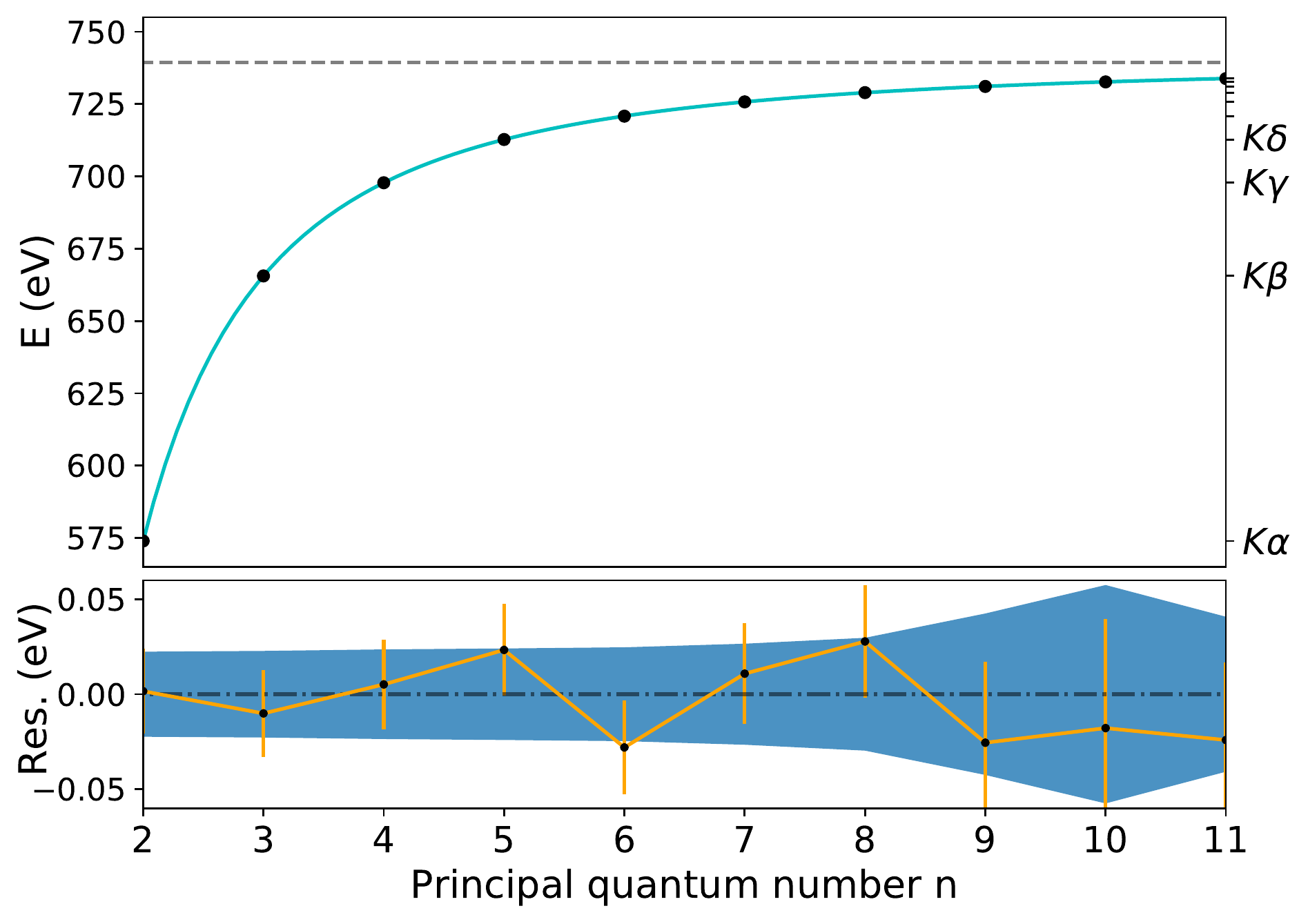}
		\caption{Fit of the Rydberg formula to the measured He-like $K$ series with nuclear effective charge Ze, quantum defect $\delta$ and ionization potential ($E_\text{IP}$) as a free parameter. The blue shaded area indicates the confidence interval.}
		\label{ionizationpotential}
	\end{center}
\end{figure}

Fitting this model (Fig.~\ref{ionizationpotential}) yields an ionization potential $E_\text{IP}$) of O$^{6+}$ of {$E_\text{IP}=739.336(16)$\,eV}, which agrees very well with the 739.326 82(6)\,eV predicted by~\citet{Drake1988} and 739.326 262\,eV by  \citet{Tupitsyn2020} (Table~\ref{tab2}).
%$\mathrm{IP}={739.327 \pm 0.016} \,\, \mathrm{eV}$

\begin{center}
\setlength{\tabcolsep}{7pt}
\renewcommand{\arraystretch}{1}
\begin{table*}[!ht]
\caption{Experimental excitation energies of He-like and Li-like oxygen absorption resonances (in eV). %The measured energy differences to the He-like $K\alpha$ are referenced to its theoretical value of 573.961058\,eV (from \citet{Yerokhin2019}) for obtaining their respective absolute energies (abs. $E$). 
Results are compared to calculations using the relativistic configuration-interaction method ({\sc FAC}: O~\textsc{vi}, this work; RCI-QED: O~\textsc{vii}~\cite{YerokhinErratum2017}; RCI: O~\textsc{vii}~\cite{Savukov2003}, O~\textsc{vi}:~\cite{Natarajan2015}), NIST: values compiled in ~\cite{NIST_ASD}, and Exp.: experimental results from \cite{Engstrom1995}}.

\label{tab1}
\vspace{1mm}
\centering
%\begin{tabular}{c c c c c c c c}

\begin{tabular}{c c l c c l l c l}
\hline \hline
Ion & Label & Final states & This Exp. & {\sc FAC} & RCI-QED  & RCI & NIST & Exp. \\ 
\hline \\
O~\textsc{vii} & $K\alpha$   & $[1s_{1/2}\,2p_{3/2}]_1$& 573.96(2) &  & 573.9614(5) &574.000 &573.94777 & 573.949(8)\\
O~\textsc{vii} & $K\beta$    &$[1s_{1/2}\,3p_{3/2}]_1$& 665.58(2) &  &  665.5743(3)& 665.615&665.61536 & 665.565(14)\\
O~\textsc{vii} &$K\gamma$   &$[1s_{1/2}\,4p_{3/2}]_1$& 697.79(2)   &  &  697.7859(3) & 697.834&697.79546 & 697.783(27)\\
O~\textsc{vii} &$K\delta$   &$[1s_{1/2}\,5p_{3/2}]_1$& 712.74(2)   &  &  712.7221(3) & 712.758&712.71696 & 712.717(82)\\
O~\textsc{vii} &$K\epsilon$ &$[1s_{1/2}\,6p_{3/2}]_1$&720.81(2) &  &  720.8434(3) & 720.880&720.83792 & \\
O~\textsc{vii} &$K\zeta$    &$[1s_{1/2}\,7p_{3/2}]_1$&725.75(3) && 725.7432(3) & &725.64727 &\\
O~\textsc{vii} &$K\eta$     &$[1s_{1/2}\,8p_{3/2}]_1$& 728.95(3)&&             &         & \\
O~\textsc{vii} &$K\theta$   &$[1s_{1/2}\,9p_{3/2}]_1$& 731.08(4)&&            &          & \\
O~\textsc{vii} &$K\iota$    &$[1s_{1/2}\,10p_{3/2}]_1$& 732.65(6)&&           &              & \\
O~\textsc{vii} &$K\kappa$   &$[1s_{1/2}\,11p_{3/2}]_1$& 733.80(4) &&           &           &  \\
\\
\hline
\\
O~\textsc{vi} & & $[1s\,2s^2]_{1/2}$ \footnote{Forbidden line.} & &  548.36 & 550.699(8) & 550.67 & \\
O~\textsc{vi} &$K\alpha$&$[(1s\,2s)_0\,2p_{3/2}]_{3/2}$& &566.81 & 567.7216(47) & \\
O~\textsc{vi} &$K\beta_1$&$[(1s\,2s)_1\,3p_{3/2}]_{3/2;1/2}$&640.20(2) & 638.50, 638.51 & &\\
O~\textsc{vi} &$K\beta_2$&$[(1s\,2s)_0\,3p_{1/2,3/2}]_{1/2;3/2}$&646.96(2) & 644.69, 644.70 & &\\
O~\textsc{vi} &$K\gamma$&$[(1s\,2s)_1\,4p_{3/2}]_{3/2;1/2}$&667.18(3) & 665.30, 665.30 & &\\
O~\textsc{vi} &$K\delta$&$[(1s\,2s)_1\,5p_{3/2}]_{3/2;1/2}$)&678.90(4) & 677.11, 677.11 & &\\
\\

\hline \hline
\end{tabular}
\end{table*}
\end{center}

For the fluorescence-photon energy calibration of the SDD we also use the $K$ series transitions of He-like oxygen (Fig.~\ref{fullview}a) up to $K\zeta$. Each component of this series shows a well-resolved, elastic single photon decay to the ground state, thus allowing us to assign to the centroids of their $y$ projections the same energies as the respective exciting photon of the $x$ projection. 

\section{OBSERVATION OF TEOP TRANSITIONS IN Li-LIKE OXYGEN}
Now we turn our attention to Li-like O$^{5+}$, a very essential astrophysical ion. Usually, inner-shell vacancies relax into the ground state by Auger decay (AD) emitting electrons, by one-electron-one-photon (OEOP) transitions, or by cascades thereof. However, TEOP processes are possible, albeit at usually slower rates than the other processes. The, customarily called, multi-electron transitions were first considered by \citet{Heisenberg1925}, while \citet{Condon1930}, and \citet{Goudsmit1931} found the pertinent selection rules. More than 40 years later,~\citet{Woelfli1975} observed TEOP x-ray photons following production of multiple inner-shell vacancies in heavy-ion-atom collisions. Later, they were seen in ion-ion collisions \cite{Briand1974,Mikkola1979,Stoller1977,Ahopelto1979,Salem1982,Salem1984,Auerhammer1988}, laser-produced plasmas \cite{Rosmej2001}, and EBIT experiments \cite{Zou2003,harman2019}. Various approaches for calculating transition rates and cross sections were introduced  \cite{Kelly1976,Aberg1976,Gavrila1978,Baptista1986,Mukherjee1988,Saha2009,Kadrekar2010,Natarajan2013,Natarajan2015}. Recently, Fano-like interference between the TEOP transition and dielectronic recombination was investigated theoretically~\cite{lyashchenko2020}. In general, the TEOP transition was regarded as second-order process that could only be noticeable when otherwise competing OEOP transitions and AD were forbidden due to either selection rules or being intra-shell radiative transitions \cite{Indelicato1997,drake1999,Zou2003,Trotsenko2007,Marques2012}. Here, in contrast, TEOP transitions suppress usually dominant allowed channels.

\subsection{Measurement of fluorescence-photon energies}
\begin{figure*}[!ht]
	\includegraphics[width=0.8\textwidth]{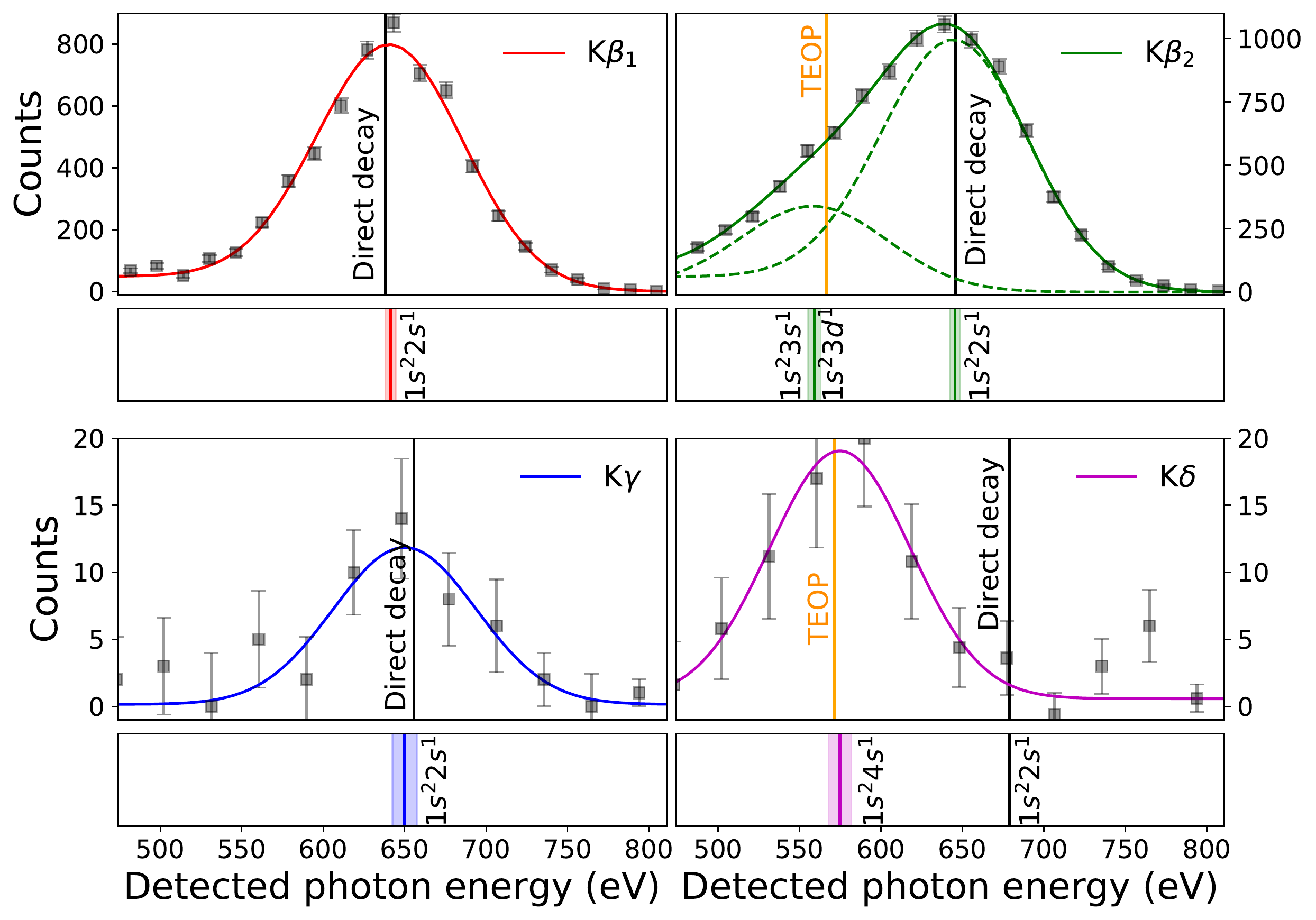}
	\caption{Fitted fluorescence decay spectra of the $K\beta_{1,2}$, $K\gamma $ and $K\delta$ absorption resonances in Li-like O$^{5+}$ ions. Lines in the same color as the fit curves mark the corresponding decay energies (labeled by their respective final configuration), and the shaded area their uncertainties. The orange line shows the theoretical TEOP channel position and the black line the theoretical direct decay channel position. Here $K\beta_{1,2}$ contain data from separate scans with higher statistics and were fitted withthe addition of the Compton lower-energy tail.}
	\label{teopspectra}
\end{figure*}

We measure the TEOP-transition energies in fluorescence to distinguish them from other channels. For both Li-like $K{\beta_1}$ at \SI{640}{\electronvolt} and $K\gamma$ at \SI{666}{\electronvolt}, we observed the OEOP radiative decay channel into the ground state: $1s\,2s\,3p\rightarrow1s^2\,2s$ and $1s\,2s\,4p\rightarrow1s^2\,2s$. 
\begin{center}
	\begin{table}[b]
		\caption{Fit results in comparison with available theoretical and experimental values.}
		\label{tab2}
		\begin{tabular}{ ccccc }
			\hline\hline
			& Fit & \cite{scofield} & \cite{Drake1988} & \cite{Tupitsyn2020}\\
			\hline
			%IP&739.327(16) & 739.30 & 739.32682(6)\\ 
			$E_\text{IP}$&739.336(16) & 739.3& 739.32682(6) & 739.326262\\ 
			$\text{Z}_\text{eff}$&7.008(3)& & &\\
			$\text{$\delta$}$&$-0.0014(9)$ & &\\
			\hline \hline
		\end{tabular}
		\label{fitresults}
	\end{table}
\end{center}
Hereafter, we refer the radiative decay channel towards the ground state as direct decay (DD), to distinguish from sequential two-photon decays (TPD), such as $1s\,2s\,3p\rightarrow1s\,2s^2\rightarrow1s^2\,2p$. Direct decay is the time-inverse process of photoexcitation (PE), and the overall process of PE plus DD is equivalent to elastic fluorescence emission, as apparent for $K{\beta_1}$ and $K\gamma$ in the decay spectrum of Fig.~\ref{fullview}a. Figure~\ref{teopspectra} shows the decay spectra for $K{\beta_1}$ and $K\gamma$, confirming these DD channels. 
However, as also displayed in Fig.~\ref{teopspectra}, the 646-eV $K{\beta_2}$ line and $K\delta$ both reveal different radiative decay channels besides the expected DD. While $K{\beta_2}$ appears to have a minor contribution to the main elastic DD channel, $K{\delta}$ shows a dominant inelastic channel and no elastic one. To understand this, we performed calculations of the main decay channels of the lines presented in Fig.~\ref{teopspectra} with the Flexible Atomic Code ({\sc FAC}) \cite{Gu2008}, which provide us with transition rates missing in the high-accuracy calculations of \citet{Yerokhin2019}.

\begin{figure*}%[t]
\includegraphics[width=0.75\textwidth]{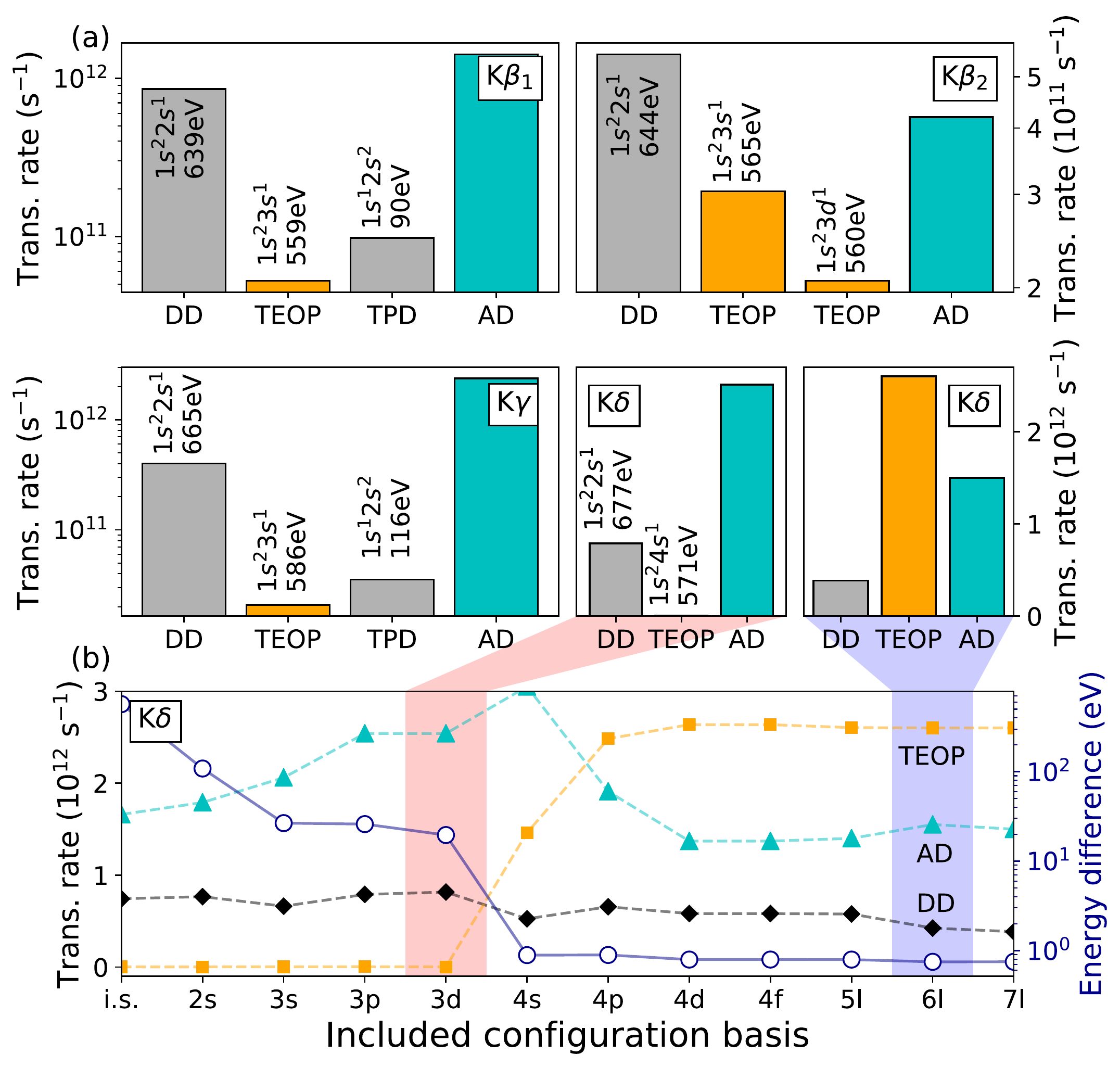}
\caption{a) {\sc FAC} transition-rate calculations for the strongest decay channels following resonant photoexcitation of the $K\beta_{1,2}$, $K{\gamma}$, and $ K{\delta}$ transitions in O$^{5+}$ ions, and b) their convergence depending on the configuration basis size (see the text). For $K{\delta}$, two stages of convergence are highlighted: red indicates incomplete and blue the final result. The respective final states and transition energies are labeled accordingly. Here DD denotes the direct one-photon decay to the ground state, TEOP the two-electron-one photon decay, TPD the two-photon decay cascade, and AD Auger decay.}
\label{lihistogram}
\end{figure*}

While doubly-excited states commonly relax by AD, our {\sc FAC}~\cite{Gu2008} calculations show that this channel is only relevant for $K{\beta_1}$ and $K{\gamma}$ (see Fig.~\ref{lihistogram}), and also confirm a main DD for $K{\beta_1}$ and $K{\gamma}$. After PE of $K{\beta_2}$, DD competes with the TEOP transition feeding into the $1s^2\,3s$ and $1s^2\,3d$ states (roughly 5\,eV apart) that can radiatively decay through various cascades. This results from configuration mixing with the $1s\,2p\,3s$ and $1s\,2p\,3d$ states. For $K\delta$, no significant DD could be observed. Here, the upper state dominantly relaxes through a TEOP transition to the $1s^2\,4s$ state.

The question is, what makes the usual DD to the ground state so weak? As shown in Fig.~\ref{fullview}c, the excited $1s\,2s\,5p$ state has a near degeneracy ($0.8$~eV) with a state of the same total angular momentum and parity, $1s\,2p\,4s$, which is also the upper state of a Li-like satellite of the He-like $K\alpha$ line. Thus, the excited states strongly mix with these, which have much higher decay rates towards $1s^2\,4s$ (on the order of \SI{E12} s$^{-1}$). This suppresses the one-photon DD to the ground state, as can be seen in Fig.~\ref{teopspectra}. 

\setlength{\tabcolsep}{35pt}
\renewcommand{\arraystretch}{1}
\begin{table}[!ht]
	\begin{center}
		\caption{Configurations added to the CI basis set for each calculation having the CI label of Fig.~\ref{cibasis}; each set keeps all the configurations of the sets above it. }
		\label{tab3}
		\vspace{3mm}
		%\centering
		%\begin{tabular}{c c c c c c c c}
		
		\begin{tabular}{c r  }
			\hline \hline
			CI  label & Configuration  set \\ 
			\hline \\
			
			initial set	&	$1s^2 2l$	\\
			&	$1s^2 4l $	\\
			&	$1s 2s 5l $	\\
			$2l$	&	$1s 2l2l'$	\\
			$3s$	&	$1s^2 3s$	\\
			&	$1s 2l 3s$	\\
			$3p$	&	$1s^2 3p$	\\
			&	$1s 2l 3p$	\\
			$3d$	&	$1s^2 3d$	\\
			&	$1s 2l 3d$	\\
			$4s$	&	$1s 2l 4s$	\\
			$4p$	&	$1s 2l 4p$	\\
			$4d$	&	$1s 2l 4d$	\\
			$4f$	&	$1s 2l 4f$	\\
			$5l$ 	&	$1s^2 5l $	\\
			$6l$	&	$1s^2 6l $	\\
			&	$1s 2l 6l' $	\\
			$7l$	&	$1s^2 7l $	\\
			&	$1s 2l 7l' $	\\
			\hline \hline
		\end{tabular}
	\end{center}
\end{table}

\begin{figure*}[t]
	\begin{center}
		\includegraphics[width=0.75\textwidth]{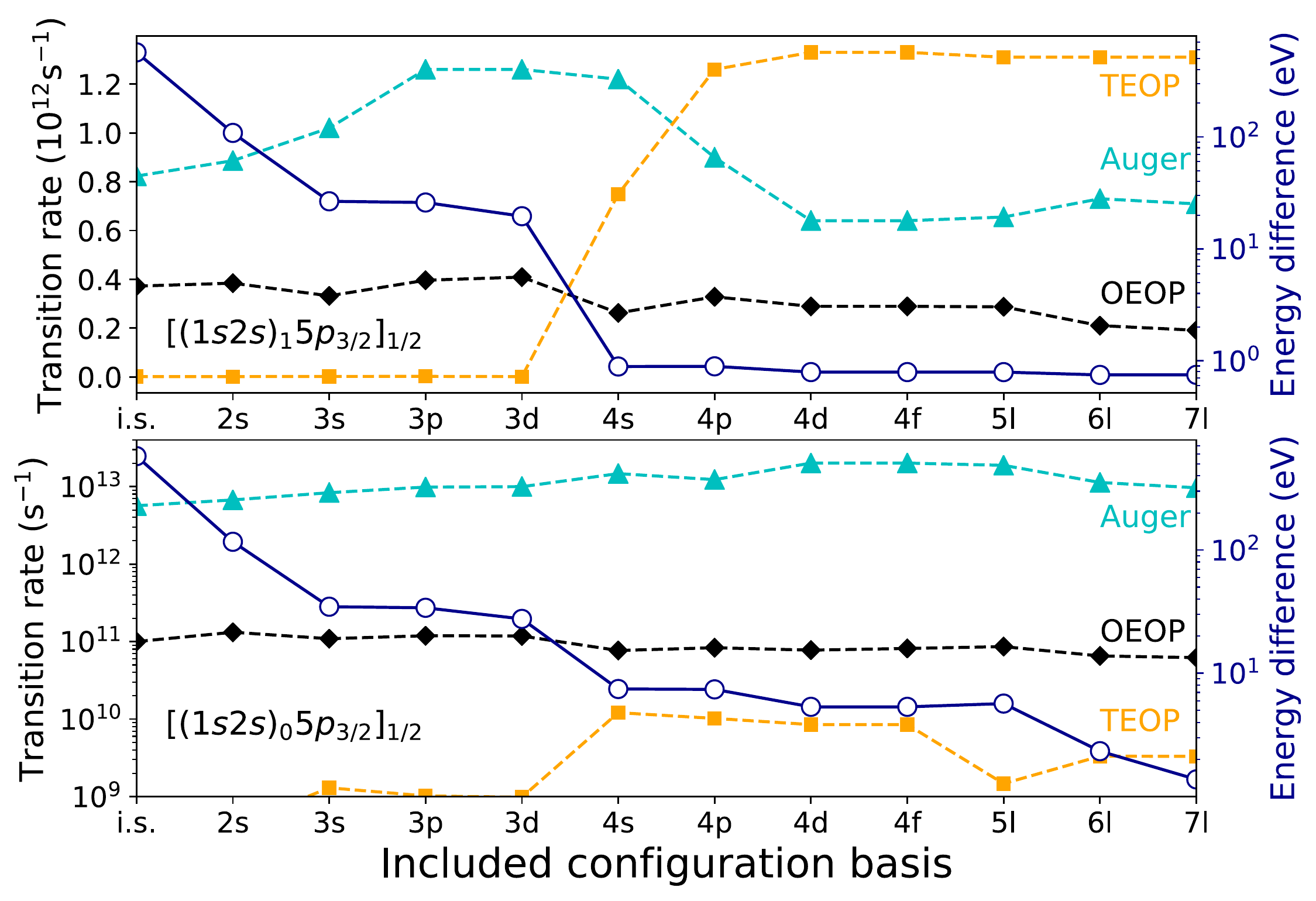}
		\caption{Calculated decay rates and energy degeneracies of the initial states $[(1s2s)_15p_{3/2}]_{1/2}$ (top) and $[(1s2s)_05p_{3/2}]_{1/2}$ (bottom) with different CI basis sets (see Table~\ref{tab3} in text). Both transition rates to final states $[1s^2 2s]_{1/2}$ (the OEOP transition) and $[1s^2 4s]_{1/2}$ (the TEOP transition) are included, along with the Auger rate to $1s^2$ state (left y-scale). Open circles represent the energy difference (degeneracy) in eV (right y-scale).}
		\label{cibasis}
	\end{center}
\end{figure*}

\setlength{\tabcolsep}{16pt}
\renewcommand{\arraystretch}{1.1}
\begin{table*}
\caption{Core-hole relaxation following $K\beta_{1,2}$, $K{\gamma} $, and $ K{\delta}$ resonant photon excitation in O$^{5+}$ ions. The first column lists the excitation transitions. Subsequent columns indicate the decay channels' respective initial and final configurations, peak photon energies (from the silicon-drift detector), final-state energies (calculated with {\sc FAC}~\cite{Gu2008}), and transition type (DD denotes direct photon decay, and TEOP the two-electron--one-photon transition). The DD-to-TEOP intensity ratio  is also given. All energies are in eV.}
\vspace{2mm}
    \begin{tabular}{ llllll }
    \hline \hline
     Line & Initial configuration  & Final configuration & Photon energy & {\sc FAC} & Type\\ 
     \hline 
    $K\beta_1$&$1s\,2s\,3p$ & $1s^2\,2s$ & 641.68(3.3)  & 638.50 & DD\\ 
    $K\beta_2$&$1s\,2s\,3p$ & $1s^2\,2s$ & 645.8(3.3)  & 644.70 & DD\\
    $K\beta_2$&$1s\,2s\,3p$ & $1s^2\,3s$ & 559.4(3.8) & 565.40 & TEOP\\ 
    			 && $1s^2\,3d$ &    blend    & 560.2 & TEOP\\ 
    $K\gamma$&$1s\,2s\,4p$ & $1s^2\,2s$ & 650.2(7.8)  & 665.30  & DD\\ 
    $K\delta$&$1s\,2s\,5p$ & $1s^2\,4s$ & 574.9(7.6)  & 570.71  & TEOP\\ 
    \hline
   branching ratio& $K\beta_2$ & DD-to-TEOP ratio & Exp.:\,1.73(19)& FAC:\,1.39 \\
    \hline \hline
    \end{tabular}

\label{tabelleabregung}
\end{table*}
%

%\subsection{Theoretical Analysis}

\subsection{Role of configuration mixing: Calculations}
We investigate the underlying quantum processes by calculating electronic energies, transition rates, and Auger rates with the relativistic configuration-interaction package {\sc FAC}~\cite{Gu2008}. The convergence of the configuration mixing was studied by varying the size of the configuration-interaction (CI) basis set, as listed in Table~\ref{tab3}. This allowed us to identify the key configurations leading to a strong TEOP rate. 

Figure~\ref{cibasis} displays the effect of the CI basis size on the transition rates from the upper state $[(1s2s)_1 5p_{3/2}]_{1/2}$ towards the final states $[1s^2 2s]_{1/2}$ (the OEOP transition) and $[1s^2 4s]_{1/2}$ (the TEOP transition), and on the Auger rate to the $1s^2$ state (the only possible Auger channel). All the decay rates have an allowed electric dipole contribution that dominates higher-order multipoles. The energy degeneracy, defined as the smallest energy difference between this initial state and the nearest one having the same total momentum and parity symmetry in a different configuration, is also represented. While the OEOP rate is nearly independent of the CI-basis set, a sudden increase of three orders of magnitude in the TEOP transition appears when the $1s 2p 4s$ configuration is included, which leads to a mixed state with contributions of order of 75\% from $[(1s2s)_1 5p_{1/2,3/2}]_{1/2}$ and of order of 25\% from $[(1s 2p_{1/2,3/2})_1 4s]_{1/2}$. Adding further configurations lets the TEOP rate converge towards a value that is approximately twice that of the Auger process, and eight times that of OEOP decay.
The above-mentioned energy degeneracy becomes more pronounced with the inclusion of further configurations, as the energy separation decreases from tens of eV to order of 0.8~eV. This causes the mixing coefficient to grow by order of 28\% for $[(1s2p_{1/2,3/2})_1 4s]_{1/2}$, which combined with the high decay rate of the $[1s 2p 4s]_{1/2}\rightarrow [1s^2 4s]_{1/2}$ transition (the satellite of the He-like $K\alpha$) makes the TEOP rate for $[1s 2s 5p]$ predominant. The initial $[(1s2s)_1 5p_{3/2}]_{3/2}$ state also follows a similar behavior. The Auger rates drop due to mixing with configurations of higher orbital momentum having lower Auger rates. The TEOP rate for the initial state $[(1s2s)_0 5p_{1/2}]_{1/2}$ does not show a significant increase with inclusion of $[1s 2p 4s]_{1/2}$ in the basis set. This is due to the energy difference with  $[1s^2 4s]_{1/2}$ being of order of a few eV, which reduces the mixing coefficient of the $[1s 2p 4s]_{1/2}$ configuration to 1.3\%.

%Summarizing our theoretical analysis, Fig.~\ref{lihistogram} shows that including the near degenerate $1s\,2p\,4s\,$ state in the calculation of the $K_\delta$ decay rates increases the TEOP rate by more than 3 orders of magnitude. With this strong configuration mixing, it becomes also plausible that a TEOP-PE process from the ground state towards $1s\,2p\,4s\,$ can further enhance inelastic decay. However, a comparison of the OEOP ($1s^2 \,2s\,\rightarrow$ $1s\,2s\,5p\,$) and TEOP ($1s^2 \,2s\,\rightarrow$ $1s\,2p\,4s\,$) PE rates, which are proportional to the respective decay rates shown in Fig.~\ref{lihistogram}\,(c), show that the rate for TEOP-PE ($\approx 2\times10^{10}~\mbox{s}^{-1}$) is one order of magnitude lower than the one for OEOP-PE ($\approx 1\times10^{11}~\mbox{s}^{-1}$).
%
% Change by CS on 23rd Sept
Summarizing our theoretical analysis, Fig.~\ref{lihistogram} shows that including the near degenerate $1s\,2p\,4s\,$ state in the calculation of the $K\delta$ decay rates increases the TEOP rate by more than three orders of magnitude. 
Besides this, we also checked another plausible photoexcitation channel $1s^2 \,2s\,\rightarrow$ $1s\,2p\,4s\,$ that can also decay to $1s^2 \, 4s$ ground state. This alternative path can possibly further enhance the observed TEOP channel. 
However, a comparison of photoexcitation rates between $1s^2 \,2s\,\rightarrow$ $1s\,2p\,4s\,$ (approximately equal to $2\times10^{10}~\mbox{s}^{-1}$) and $1s^2 \,2s\,\rightarrow$ $1s\,2s\,5p\,$ (approximately equal to $1\times10^{11}~\mbox{s}^{-1}$) shows an order of magnitude difference between them. 
Therefore, we emphasize that the $1s\,2p\,4s\,$ state can only be populated via $K\delta$ photoexcitation and decay via the TEOP channel, as observed in the present experiment.

\subsection{Determination of the ratio DD/TEOP}
Our measured $K{\beta_2}$ decay energies are listed in Table~\ref{tabelleabregung}. Because the respective fluorescence-decay  channels can be barely resolved by the SDD (cf. Fig.~\ref{teopspectra}), their centroids were only determined with an uncertainty in the 0.5--1.3\% range; this can be improved in the future, e.~g., by using a high-resolution x-ray microcalorimeter \cite{Durkin2019}. Since measuring line ratios is an essential plasma diagnostic tool, and stringently tests theory, we extract the ratio of the DD rate to the TEOP transition rate from the measured spectra (Fig.~\ref{teopspectra}).
While the DD-to-TEOP ratio of $K\delta$ could not be accurately determined due to low statistics in the DD channel, it was nonetheless possible to quantify the DD-to-TEOP ratio for the Li-like $K\beta_2$ emission. For this we characterized the x-ray detector taking into account the transmission of the 500-nm Al filter placed in front of it to block visible and UV radiation, and also the low-energy Compton tail of the detector response line shape. This tail was obtained from a fit to the $K\beta_1$ transition, which should only have the elastic channel, as configuration mixing with other states is small, and used as fit function for the DD channel of $K\beta_2$. We estimate the uncertainty of the Al-filter transmission from its thickness ($500\pm100$~nm), adding a contribution of $\pm 15\%$ to the ratio-error budget. The soft x-ray spectral sensitivity of the SDD strongly depends on the thickness of native silicon dioxide layer on its surface, which is not well known (20 to 50 nm), on the top-electrode materials, and on the slow condensation of water on its cold surface during the experiment. Since one of the decay channels (the TEOP transition) of $K\beta_2$ is close to the oxygen $K$ edge, these layers can significantly change the ratio of the transmission coefficients for the $1s2s3p\rightarrow1s^2$ transition at 644\,eV versus the one of the $1s2s3p\rightarrow1s^23d$ at 565\,eV. With these caveats, and assuming a filter transmission ratio (644\,eV-to-565\,eV) approximately equal to $ 1.25\pm0.2$ and a SDD sensitivity ratio (644\,eV-to-565\,eV) of approximately $ 1.15\pm0.5$, the observed intensity ratio of 2.93 yields a ratio of approximately $2.0\pm0.8$, which is compatible with the {\sc FAC} prediction.

\setlength{\tabcolsep}{4pt}
\renewcommand{\arraystretch}{1}
\begin{table}[t]
\caption{Decay rates for the $K\beta_2$ states $[(1s 2s)_0 3p_{1/2}]_{1/2}$ ($J=1/2$) and $[(1s 2s)_0 3p_{3/2}]_{3/2}$ ($J=3/2$). Values are in s$^{-1}$.}
\label{tab5}
%\vspace{2mm}
\centering
%\begin{tabular}{}
\begin{tabular}{lccccc  }
\hline 
\hline
State& DD rate &   \multicolumn{3}{c}{TEOP rate} \\
\cline{3-5}
 \phantom{$\begin{array}{c}\\ \frac{1}{2} \rightarrow 1/2 \end{array}$} &  $1s^2 2s$ & $1s^2  3s$ & $1s^2  3d_{3/2}$ & $1s^2  3d_{5/2}$  \\
\hline \\
%\vspace{1.6mm}
$J=1/2$  &	$3.24\times 10^{11}$ & $1.17\times 10^{11}$ & $1.03\times 10^{11}$  \\
$J=3/2$  &	$3.21\times 10^{11}$ & $1.29\times 10^{11}$ & $1.02\times 10^{10}$ & $9.25\times 10^{10}$ \\

%\vspace{0.3mm}
\hline \hline
\end{tabular}
\end{table}
In a second approach, we determine the DD-to-TEOP ratio by comparing the intensity ratio K$\alpha$/K$\beta = 2.06\pm0.05$ of the simultaneously observed He-like transitions with the ratio of their theoretical Einstein coefficients, $A_{ik}/A_{ik'} = 3.53$ according to the NIST database \cite{NIST_ASD}. We normalize the observed intensities to the excitation-photon flux, and obtain a sensitivity ratio (665.61\,eV-to-573.94\,eV) approximately equal to $ 1.70\pm 0.17$, whereby the decay channels very closely match the energies of the Li-like transitions under investigation. This takes into account all previously mentioned effects of the filter and detector efficiency. When we interpolate this sensitivity ratio to the close-by Li-like transitions, the observed intensity ratio (approximately equal to 2.93) for the Li-like K$\beta_2$ decay channels results in a ratio of approximately {1.73}$\pm${0.19}, in fair agreement with our {\sc FAC} calculation.

Emission of $K\beta_2$ follows PE of the states $[(1s 2s)_0 3p_{1/2}]_{1/2}$ and $[(1s 2s)_0 3p_{3/2}]_{3/2}$ feeding the radiative decay channels listed in Table~\ref{tab5}. Their respective strengths yield a DD-to-TEOP ratio of 1.39 for an observation angle of $90^{\circ}$ (see the Appendix) and 1.43 for the solid-angle integrated total emission. Note that the decay rates from these states to the ground state are similar and likewise the corresponding PE cross sections. This cancels the effect of state population on the ratio.

\setlength{\tabcolsep}{1pt}
\renewcommand{\arraystretch}{1}
\begin{table}[t]
	\caption{Angular coefficients $\beta_{J_d \rightarrow J_f}$ for angular momenta $J_d$ and $J_f$.}
	\label{tab4}
	\vspace{2mm}
	\centering
	%\begin{tabular}{}
	\begin{tabular}{lcccccc  }
		\hline 
		\hline
		%~ \vspace{1.3mm}
		${J_d \rightarrow J_f}$
		\phantom{$\begin{array}{c}\\ \frac{1}{2} \rightarrow 1/2 \end{array}$} & \hspace{0.3cm} $ \frac{1}{2} \rightarrow \frac{1}{2} $ & \hspace{0.3cm}$\frac{1}{2}\rightarrow \frac{3}{2}$ & \hspace{0.3cm}$\frac{3}{2}\rightarrow \frac{1}{2}$ & \hspace{0.3cm}$\frac{3}{2}\rightarrow \frac{3}{2}$ & \hspace{0.3cm}$\frac{3}{2}\rightarrow \frac{5}{2}$\\ 
		\hline \\
		\vspace{1.6mm}
		$\beta_{J_d \rightarrow J_f}$ 	&	0 & 0  & $-\frac{1}{2}$ & $\frac{2}{5}$ & $-\frac{1}{10}$	\\
		%\vspace{0.3mm}
		\hline \hline
	\end{tabular}
\end{table}

\section{Conclusion}

We have demonstrated how a complex process that is difficult to disentangle in astrophysical plasmas can be isolated and studied in detail by high-resolution photon excitation. Unexpectedly strong TEOP transitions were found in an essential species, the relatively simple Li-like O$^{5+}$, showing, among other observations, evidence that the upper state of the $K\delta$ line in Li-like O$^{5+}$ mainly decays as a satellite of He-like O$^{6+}$ $K\alpha$. This produces a problematic blend in a key feature for the diagnostics of photoionized plasmas (e.g.~\cite{Mehdipour2015}). Although a strong suppression of the direct photo decay by TEOP transitions was observed in just one of several lines in Li-like oxygen, it is not far-fetched to assume that TEOP-dominated relaxation also happens in other multiply excited, multi-electron systems, and thus its contribution should not be neglected in accurate astrophysical plasma models. 

Systems with more than three electrons have richer overlapping excitations with manifold decay channels not only cause similar blends in emission and absorption spectra, but also affect the ionization balance of plasmas. The three-electron system studied here is more tractable by current theory and has allowed us to stringently test the underlying electronic correlations. This is of great importance for the diagnostics of hot gas in astrophysics. The upcoming launches of \textit{XRISM}~\cite{XRISM} and \textit{Athena}~\cite{ATHENA} urgently call for studying the position and strength of TEOP transitions that can cause shifts, or broaden the strong diagnostically important O-$K$ and Fe-$L$ lines in the 15--23~\AA~range, which are crucial for determining gas-outflow velocities of warm absorbers and density diagnostics of photoionized plasmas~\cite{drake1999,Behar2003,Schmidt2004,Gu2005,Mehdipour2015}, and needed for accurately modeling the x-ray continuum flux.

%\section{Acknowledgements}
    \begin{acknowledgments}
        Financial support was provided by the Max-Planck-Ge\-sell\-schaft and Bun\-des\-mi\-ni\-ste\-ri\-um f{\"u}r Bildung und Forschung through Project No.~05K13SJ2.
        We acknowledge DESY (Hamburg, Germany), a member of the Helmholtz Association HGF, for the provision of experimental facilities. Parts of this research were carried out at PETRAIII. 
        Work by C.S.~was supported by the Deutsche For\-schungs\-ge\-mein\-schaft Project No.~266229290 and by an appointment to the NASA Postdoctoral Program at the NASA Goddard Space Flight Center, administered by Universities Space Research Association under contract with NASA.
        P.A.~acknowledges support from Funda\c{c}\~{a}o para a Ci\^{e}ncia e a Tec\-no\-lo\-gia (FCT), Portugal, under Grant No.~UID/FIS/04559/2020(LIBPhys) and under Contract No.~SFRH/BPD/92329/2013. 
        Work at UNIST was supported by the National Research Foundation of Korea (Grant No. NRF-2016R1A5A1013277).
        Work at LLNL was performed under the auspices of the U.~S.~ Department of Energy under Contract No.~DE-AC52-07NA27344 and supported by NASA grants to LLNL. 
        M.A.L.~and F.S.P.~acknowledge support from NASA’s Astrophysics Program. 
       \end{acknowledgments}

\appendix
\section{ANGULAR DISTRIBUTION}

In our experiment, fluorescence photons were observed at $90^{\circ}$, so we took into account the angular distribution pattern of their emission in the experimental determination of the DD-to-TEOP ratio. We treated PE and subsequent radiative decay as a two-step process within the $E1$ dipole approximation, which is appropriate as the main multipole channel of both the TEOP transition and DD transition is of $E1$ type. We assume that the ground state is not initially aligned, allowing us to apply for the angular distribution the formula given by \citet{Balashov2013},
\begin{equation}\nonumber
W(\theta)=\frac{W_0}{4\pi}\left( 1 -\frac{\beta_{J_d \rightarrow J_f}}{2} P_2(\cos{\theta})\right).
\end{equation}
 This formula is valid for circularly polarized incident photons, as in the present experiment, and yields a dependence of the radial angle $\theta$ ($z$ axis or quantization axis alongside the incident photon beam propagation axis and magnetic field) in terms of a second rank Legendre polynomial $P_2$, with $W_0$ being the total emission. The angular coefficients $\beta_{J_d \rightarrow J_f}$ from the photoexcited state with total angular momentum $J_d$ to various final states of interest for the case $K\beta_2$ are given in Table~\ref{tab4}.

\end{document}